\documentclass[nonacm,sigplan]{acmart}

\usepackage{subfig}
\usepackage{stfloats}
\usepackage{enumitem}

\usepackage{url}
\usepackage[normalem]{ulem}

\usepackage{amsmath,amssymb,amsfonts}
\usepackage{algorithmic}
\usepackage{multirow}
\usepackage{stfloats}

\usepackage{amsthm}
\usepackage{braket}
\usepackage{tcolorbox}

\usepackage{etoolbox}
\AtBeginEnvironment{tabular}{\footnotesize}

\definecolor{aliceblue}{rgb}{0.94, 0.97, 1.0}
\usepackage{xcolor}
\usepackage{tcolorbox}
\tcbset{width=1\columnwidth,boxrule=1pt,colback=aliceblue,arc=1pt,left=2pt,right=2pt,boxsep=0.5pt}

\usepackage[flushleft]{threeparttable}

\newcommand{\ignore}[1]{}

\usepackage{url}

\begin{document}

\title{
  Skipper: Improving the Reach and Fidelity of Quantum Annealers by Skipping Long Chains
  }

\author{Ramin Ayanzadeh} 
\affiliation{%
  \institution{Georgia Institute of Technology}
  \city{Atlanta}
  \country{USA}
}

\author{Moinuddin Qureshi} 
\affiliation{%
  \institution{Georgia Institute of Technology}
  \city{Atlanta}
  \country{USA}
}

\begin{abstract}

Quantum Annealers (QAs) operate as single-instruction machines, lacking a SWAP operation to overcome limited qubit connectivity. 
Consequently, multiple physical qubits are \emph{chained} to form a program qubit with higher connectivity, resulting in a drastically diminished effective QA capacity by up to 33x.
We observe that in QAs: (a) chain lengths exhibit a power-law distribution, a few \emph{dominant chains} holding substantially more qubits than others; and (b) about 25\% of physical qubits remain unused, getting isolated between these chains.
We propose \emph{Skipper}, a software technique that enhances the capacity and fidelity of QAs by skipping dominant chains and substituting their program qubit with two readout results.
Using a 5761-qubit QA, we demonstrate that Skipper can tackle up to 59\% (Avg. 28\%) larger problems when eleven chains are skipped. 
Additionally, Skipper can improve QA fidelity by up to 44\% (Avg. 33\%) when cutting five chains (32 runs).
Users can specify up to eleven chain cuts in Skipper, necessitating about 2,000 distinct quantum executable runs. 
To mitigate this, we introduce \emph{Skipper-G}, a greedy scheme that skips sub-problems less likely to hold the global optimum, executing a maximum of 23 quantum executables with eleven chain trims.
Skipper-G can boost QA fidelity by up to 41\% (Avg. 29\%) when cutting five chains (11 runs).
        
\end{abstract}

\maketitle 
\pagestyle{plain} 


\section{Introduction}

Quantum computers (QCs) have the potential to solve certain problems beyond the capabilities of classical computers~\cite{arute2019quantum,villalonga2020establishing,preskillNISQ,king2021scaling,wu2021strong}. 
Two main types of QCs exist: 
digital machines, exemplified by industry leaders like IBM~\cite{IBMQ}, Google~\cite{GoogleAI}, IonQ~\cite{IonQ}, and Quantinuum~\cite{quantinuum},  
and analog devices such as superconducting \emph{Quantum Annealers} (\emph{QAs}) developed by D-Wave~\cite{D-Wave}, as well as neutral atom platforms by QuEra~\cite{QuEra} and PASQAL~\cite{PASQAL}. 
Both digital and analog QCs have polynomial equivalent computing power~\cite{aharonov2008adiabatic,albash2018adiabatic}.
For instance, QAs have demonstrated their potential in tackling real-world applications such as finance~\cite{elsokkary2017financial}, drug discovery~\cite{mulligan2020designing}, cryptography~\cite{peng2019factoring,hu2020quantum}, Boolean Satisfiability (SAT)~\cite{su2016quantum,ayanzadeh2020reinforcement,ayanzadeh2018solving,ayanzadeh2019sat}, planning and scheduling~\cite{inoue2021traffic,rieffel2015case,venturelli2015quantum,tran2016hybrid}, linear algebra~\cite{o2018nonnegative}, and signal processing~\cite{ayanzadeh2019quantum,ayanzadeh2020ensemble}, 
extending beyond application-specific acceleration.

\begin{figure}[htb]
    \centering
      \includegraphics[width=\columnwidth]{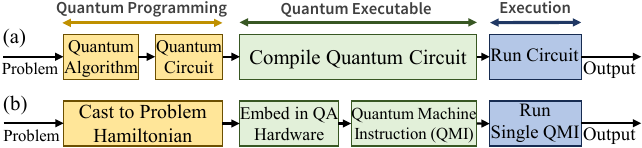}
    \caption{
        Operation Model of QCs: 
        (a) Digital QCs execute compiled quantum circuits.  
        (b) Analog QAs execute the embedded problem Hamiltonian. 
        \emph{Operating as single-instruction machines, analog QAs do not incorporate quantum circuits.} 
}       
    \label{fig:QC_operation_models}
\end{figure}

While both QC types are accessed via the cloud~\cite{AmazonBraKet,MicrosoftAzure,D-Wave}, their operation models and design trade-offs differ significantly~\cite{ayanzadeh2022equal}.
In digital QCs (namely the gate-based or circuit model quantum computing), as shown in Fig.~\ref{fig:QC_operation_models}(a), qubits undergo a scheduled sequence of quantum operations defined by the quantum algorithm to directly manipulate their states~\cite{nielsen2010quantum}.
Conversely, as shown in Fig.~\ref{fig:QC_operation_models}(b), analog QCs operate as single-instruction systems, where the qubit environment is incrementally modified based on the evolution of a physical system, called \emph{Hamiltonian}, thereby allowing natural qubit evolution and indirect state alteration~\cite{ayanzadeh2022equal,albash2018adiabatic,mcgeoch2020theory}.

\begin{figure}[htb]
    \captionsetup[subfigure]{position=top} 
    \centering
    \subfloat[]{
        \includegraphics[width=0.49\columnwidth]{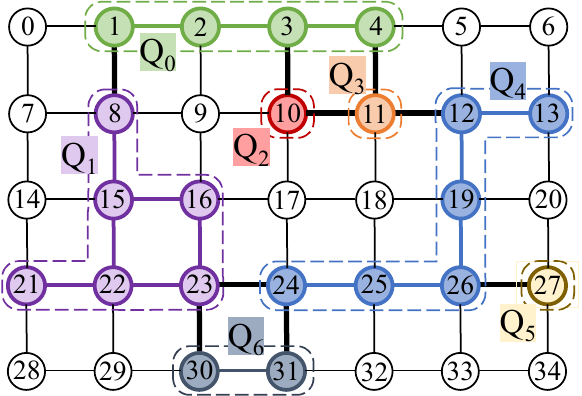}
    }
    \subfloat[]{
        \includegraphics[width=0.24\textwidth]{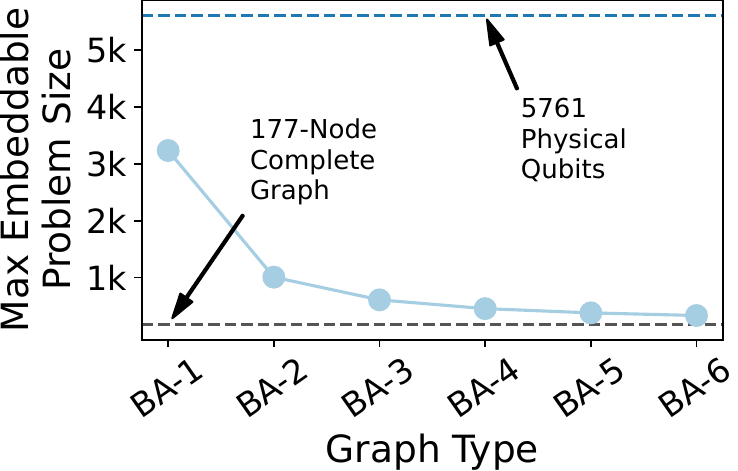}
    }
    \caption{
(a) Embedding seven program qubits $(Q_i)$ onto a $5\times7$ grid of physical qubits.
(b) Max embeddable Barabasi--Albert (BA) graphs on a 5761-qubit QA device for different preferential attachment factors (${m}$), ranging from sparse BA-1 ($m=1$) to highly dense BA-6 ($m=6$) structures.
            }    
    \label{fig:intro_embedding}    
\end{figure}

Full connectivity of qubits at scale is infeasible. 
In digital QCs, compilers introduce SWAP operations to make physical qubits adjacent~\cite{zulehner2018efficient,murali2019noise,tannu2019not}. 
Conversely, analog QCs cannot apply operations to qubits, thus preventing the use of SWAPs for qubit routing.
Instead, QAs employ \emph{embedding}~\cite{zbinden2020embedding,pelofske20234,pelofske2019solving,pelofske2022solving,barbosa2021optimizing} 
where multiple physical qubits are \emph{chained} (or entangled) to represent a program qubit with higher connectivity, as shown in Fig.~\ref{fig:intro_embedding}(a).  
Compiling quantum circuits in digital QCs preserves qubit utilization (1-to-1 mapping between program and physical qubits), however, embedding in QAs can substantially increase physical qubit utilization ~\cite{ayanzadeh2022equal}. 
For instance, the 5761-qubit QA can accommodate up to 177 program qubits with all-to-all connectivity, highlighting nearly 33x reduced \emph{logical capacity}.

Given that the hardware graph remains fixed after fabrication, QAs’ logical capacity is primarily determined by the topology of the problem graph.
Real-world applications typically involve irregular ``Power-Law’’ graphs~\cite{agler2016microbial,clauset2016colorado,gamermann2019comprehensive,goh2002classification,house2015testing,mislove2007measurement,pastor2015epidemic}, 
and Barabasi--Albert (BA) graphs are widely considered representative of such real-world graphs~\cite{albert2005scale,barabasi1999emergence,barabasi2000scale,gray2018super,kim2022sparsity,lusseau2003emergent,wang2019complex,zadorozhnyi2012structural,zbinden2020embedding}.
Figure~\ref{fig:intro_embedding}(b) illustrates the largest embeddable BA graphs on a 5761-qubit QA, ranging from sparse (with attachment factor $m=1$, BA-1) to nearly fully connected (with $m=6$, BA-6) structures.
As $m$ increases linearly, the logical capacity experiences a superpolynomial reduction, converging to the 177-node fully connected graph.

We observe that chain lengths in QAs follow a ``Power-Law’’ distribution, where a few \emph{dominant chains} are significantly longer than most other chains (see section~\ref{subsec:method_insights} for more information). 
Moreover, we observe that a significant portion of physical qubits, nearly 25\%, remain unused as they become trapped in long chains. 
Furthermore, we observe that long chains can reduce the fidelity of QAs too. 
The qubits within a chain might take different values post-measurement, called \emph{broken chains}. 
Broken chains can negatively impact QAs' reliability, and longer chains are more likely to break.

In this study, we aim to improve the capacity and fidelity of QAs through  eliminating dominant chains, as they account for a substantial portion of qubit utilization and are the main reason for isolating physical qubits.
We propose \emph{Skipper}, which \emph{prunes} these chains by removing their corresponding program qubits and replacing them with two possible measurement outcomes: -1 and +1.
By eliminating a dominant chain, Skipper accomplishes two objectives: 
(a) releasing physical qubits previously used within pruned chains, 
and (b) releasing all qubits previously trapped with the pruned chain. 
This can enable us to use all released physical qubits to accommodate more program qubits. 

However, identifying and pruning dominant chains is nontrivial. 
Chains are formed post-embedding. 
First, when a (long) chain is eliminated, the remaining embedding is likely not to be optimum, necessitating re-embedding the new problem to maximize the reliability of QAs. 
Embedding itself is nontrivial, as it can take several hours for problems at scale. 
Moreover, embedding techniques are heuristic, and they may fail to find an embedding successfully for a problem, requiring multiple attempts. 
Second, pruning the longest chain can change the position of the second-longest chain when re-embedding the problem, necessitating an embedding for every pruned chain. 
To this end, Skipper adopts a greedy approach to prune $c$ chains by sorting program qubits based on their degree and removing the top $c$ qubits simultaneously. 
We observe that this greedy approach exhibits desirable, near-optimal behavior for $c \ge 5$ chain cuts.

Importantly, the number of chain cuts in Skipper is user-defined; the system allows for a maximum of eleven chains to be cut, and this does not scale with the problem size, offering flexibility within the user's budget constraints. 
Each chain cut bifurcates the search space of the initial problem; therefore, trimming eleven chains can lead to up to 2048 sub-problems, and Skipper examines all of them to ensure guaranteed recovery.
Our experiments with a 5761-qubit QA by D-Wave demonstrate that Skipper can address up to 59\% (Avg. 28.3\%) larger problems when up to eleven chains are trimmed. 
Additionally, Skipper can significantly enhance QA fidelity by up to 44.4\% (Avg. 33.1\%), when trimming up to five chains and running 32 quantum executables.

Skipper is inspired by FrozenQubits~\cite{ayanzadeh2023frozenqubits}.
Skipper enhances both the capacity and fidelity of analog QAs. 
However, FrozenQubits has a negligible impact on the capacity of digital QCs, where one program qubit is represented with one physical qubit. 
Furthermore, FrozenQubits' performance diminishes as graph density increases, whereas Skipper effectively handles graphs ranging from sparse to dense structures.


The quantum cost of Skipper can present affordability challenges for certain users. 
We introduce \emph{Skipper-G}, a greedy approach that bypasses sub-spaces less likely to include the global optimum. 
Consequently, it runs at most 23 quantum executables, compared to the 2048 required by Skipper for trimming up to eleven chains.
It is worth noting that Skipper-G is proposed to improve QA fidelity, with its effect on increasing capacity being negligible.
Our experiments demonstrate that Skipper-G can boost QA fidelity by up to 40.8\% (Avg. 29.2\%), with five chain cuts and 11 runs.

\vspace{0.1 in}
Overall, this paper makes the following contributions:
\vspace{0.05 in}

\begin{enumerate}[ leftmargin=0.5cm,itemindent=0.cm,labelwidth=.5cm,labelsep=0cm,align=left, itemsep=0.2 cm, listparindent=0.5cm]

\item 
We show that in QAs, the chain length exhibits a ``Power-Law’’ distribution, with a few dominant chains having significantly more qubits. 
Moreover, we demonstrate that approximately 25\% of physical qubits remain unused as they become trapped within long chains.

\item 
We introduce \emph{Skipper} that enhances the capacity and reliability of QAs by cutting dominant chains, 
thereby addressing up to 59\% (Avg. 28.3\%) larger problems and improving QA fidelity by up to 44.4\% (Avg. 33.1\%), when up to eleven and five chains are cut, respectively.
To our knowledge, Skipper is the first proposal to simultaneously enhance both the capacity and fidelity of QAs.
 
\item 
We demonstrate that the quantum cost of Skipper in enhancing QA fidelity can be substantially reduced (to only 23 runs, compared to over 2000 runs in Skipper) by bypassing sub-spaces unlikely to contain optimal solutions.

\item 
We propose \emph{Skipper-G}, a greedy scheme that enhances QA fidelity by up to 40.8\% (Avg. 29.2\%), with five chain cuts and only 11 runs (compared to 32 runs in Skipper). 

\end{enumerate}

\ignore{

In addition to logical capacity, chaining can significantly impact the reliability of QAs. 
For instance, the qubits within a chain might take different values post-measurement, a phenomenon known as a \emph{broken chain}, which can negatively impact QAs' performance. 
Additionally, the distribution of chain lengths can influence QA outcomes, and uniformity in the value of chains is desirable~\cite{boothby2016fast,venturelli2015quantum}.

}

\clearpage

\section{Background and Motivation}

\subsection{Quantum Computers: Digital vs. Analog}


QCs fall into two categories: digital and analog. 
Digital QCs, like those from IBM and Google, apply precise quantum operations---defined by the quantum algorithm---to qubits in order to directly manipulate their state~\cite{nielsen2010quantum}. 
Conversely, analog QCs, like those from D-Wave and QuEra, do not directly manipulate the state of qubits. 
Instead, they apply precise changes---defined by the quantum program---to the environment in which the qubits reside, allowing the qubits to evolve and change their states naturally~\cite{albash2018adiabatic,ayanzadeh2022equal}.

\subsection{Quantum Annealers: Analog Quantum Accelerators}

Quantum annealing is a meta-heuristic for tackling optimization problems that runs on classical computers. 
\emph{Quantum Annealers} (\emph{QAs}) are a form of analog QCs that can sample from the ground state (the configuration with the lowest energy value) of a physical system, called Hamiltonian~\cite{albash2018adiabatic,ayanzadeh2021multi,ayanzadeh2022equal}. 
QAs by D-Wave are single-instruction optimization accelerators that can only sample from the ground state of the following problem Hamiltonian (or Ising model): 
\begin{equation}
    \mathcal{H}_p := \sum_{i}{\mathbf{h}_i \mathbf{z}_i} + \sum_{I \neq j}{J_{ij}\mathbf{z}_i \mathbf{z}_j}
    \label{eq:QA_H_p}
\end{equation}
acting on spin variables $\mathbf{z}_i \in {-1, +1}$, where $\mathbf{h}_i \in \mathbb{R}$  and $J_{ij} \in \mathbb{R}$ 
are linear and quadratic coefficients, respectively~\cite{ayanzadeh2022equal}.

\subsection{Operation Model of Single-Instruction QAs}

QAs operate as single-instruction computers, and during each execution trial, they only draw a single sample to approximate the global minimum of~\eqref{eq:QA_H_p}.
Therefore, we \emph{cast} real-world problems into Hamiltonians, where $\mathbf{h}$ and ${J}$ are defined in such a way that its global minimum represents the optimal solution to the problem at hand~\cite{albash2018adiabatic,ayanzadeh2022equal}.
The abstract problem Hamiltonian is then \emph{embedded} into the connectivity map of the QA hardware to generate an executable Quantum Machine Instruction (QMI)~\cite{minorminerGithub,cai2014practical}. 
Casting and embedding in QAs are akin to designing and compiling quantum circuits in digital QCs, respectively (Fig.~\ref{fig:QC_operation_models}). 
The QMI is executed for several trials, and the outcome with the lowest objective value is deemed as the ultimate result~\cite{ayanzadeh2022equal}.

\subsection{Anneal Time: Current Technological Barriers}

As the energy gap between the global minimum and the adjacent higher state diminishes linearly, the required annealing time for successful adiabaticity grows exponentially~\cite{albash2018adiabatic,das2008colloquium}, surpassing the limits of contemporary QAs~\cite{yan2022analytical}. 
Nonetheless, QAs, akin to other QCs, are advancing; subsequent generations are expected to bypass present technological constraints. 
Specifically, incorporating $XX$ terms into the time-dependent Hamiltonian can ebb the annealing time scaling from exponential to linear~\cite{nishimori2017exponential}.

\begin{figure}[t]
    \centering
    \includegraphics[width=1\columnwidth]{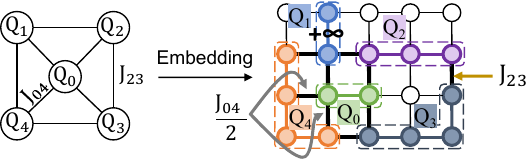}
    \caption{
        Embedding example.
}       
    \label{fig:background_embedding}
\end{figure}

\subsection{Embedding for QAs} \label{sec:embedding}    

The connectivity of QA qubits is sparse, thereby limiting users to only specify $J_{ij}$ for those qubits that are physically connected. 
Thus, the abstract problem Hamiltonian is \emph{embedded} into QA hardware where a program qubit ($Q_i$) with higher connectivity is represented by multiple physical qubits ($q_i$) called \emph{chain} (Fig.~\ref{fig:background_embedding}). 
Satisfying the following conditions is sufficient to guarantee that both the abstract Hamiltonian and the embedded Hamiltonian executed on the QA hardware have identical ground states:

\begin{enumerate}[ leftmargin=0.5cm,itemindent=0.cm,labelwidth=.5cm,labelsep=0cm,align=left, itemsep=0.2 cm, listparindent=0.5cm]

    \item 
All chains representing program qubits must be a connected component graph---i.e., there must be a path between any two qubits within a chain.

\item
There must be at least one connection between chains whose corresponding program qubits are connected.

\item
The quadratic coefficient $J_{ij}$ is distributed equally among the couplers connecting $Q_i$ and $Q_j$.

\item
The linear coefficient $\mathbf{h}_i$ is distributed equally among all physical qubits of the corresponding chain.

\item
Inter-chain quadratic coefficients must be large enough to guarantee that all qubits within a chain take an identical value---i.e., a very high penalty for broken chains.

\end{enumerate}

\subsection{Prior Work Limitations}

\subsubsection{Circuit Cutting in Digital QCs}

Circuit cutting techniques, namely CutQC~\cite{tang2021cutqc}, partition quantum circuits into smaller sub-circuits, enabling larger quantum circuits to be run on smaller QCs. 
However, a similar approach is infeasible in the analog quantum realm because:
(a) analog QAs do not incorporate quantum circuits to cut its wires; 
and (b) partitioning graphs by edge/node removal is nontrivial (e.g., highly dense graphs are non-partitionable).

\subsubsection{Solving Larger Problems on Smalle QAs}

Previous methods for solving larger problems on smaller QAs ~\cite{pelofske2022solving, okada2019improving} employ iterative or alternating approaches involving approximations, leading to reduced reliability as problem size increases. 
Additionally, convergence---even to a local optimum---is not guaranteed with these techniques.
Conversely, Skipper explores the entire search space comprehensively without resorting to approximations, and since it is not iterative, it does not face convergence issues.

\newpage
\subsubsection{Application-Specific Policies}

Recent studies have proposed methods for tackling larger instances in various domains, such as Boolean Satisfiability (SAT)~\cite{tan2023hyqsat}, Max-Clique~\cite{pelofske2023solving, pelofske2019solving, pelofske2022parallel,pelofske2021decomposition}, and compressive sensing with matrix uncertainty~\cite{ayanzadeh2019quantum,mousavi2019survey}. 
However, these techniques are tailored to their specific applications and cannot be easily adapted to other domains. 
In contrast, Skipper is versatile and can  be applied to any problem Hamiltonian. 
Moreover, reduction to SAT and Max-Clique often leads to a polynomial increase in program qubits, expanding the problem size.

\subsubsection{FrozenQubits}

Skipper is inspired by FrozenQubits~\cite{ayanzadeh2023frozenqubits}, with both methods aiming to eliminate high-degree program qubits. 
While the impact of FrozenQubits on addressing larger problems in digital QCs is minimal due to the one-to-one correspondence between program and physical qubits, 
Skipper, on the other hand, is capable of solving larger problems on QAs and enhancing QA fidelity. 
Moreover, unlike FrozenQubits, whose performance declines with increasing graph density, Skipper maintains effectiveness across a spectrum of graph densities, from sparse to dense structures.

\subsection{Goal of This Paper}

\begin{figure}[h]
    \captionsetup[subfigure]{position=top} 
    \centering
    \subfloat[]{
        \includegraphics[width=0.23\textwidth]{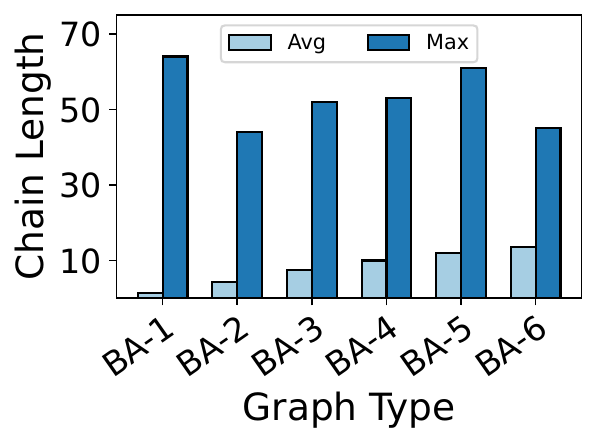}
	}
    \subfloat[]{	
        \includegraphics[width=0.24\textwidth]{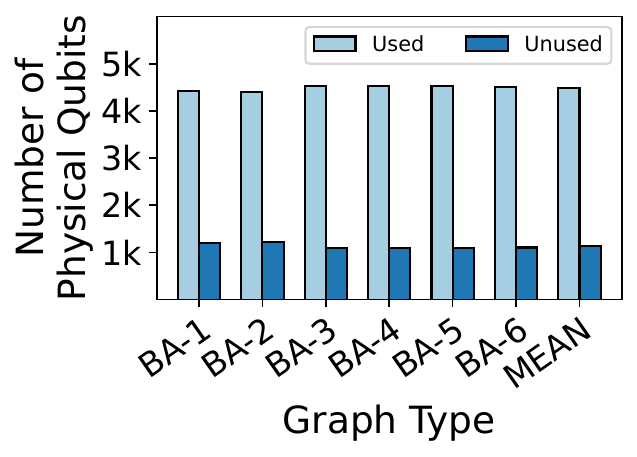}
	}
    \caption{
Maximum embeddable BA graphs on 5761-qubit QA: (a) Avg and Max chain lengths, and (b) Number of unutilized qubits. }    
    \label{fig:observation}    
\end{figure}

Figure~\ref{fig:observation}(a) shows the maximum and average chain lengths for different graph topologies embedded on a 5761-qubit QA. 
A few dominant chains contain over 7.9x as many qubits as the average chain lengths.
Furthermore, Fig.~\ref{fig:observation}(b) displays the number of unused qubits when embedding the largest possible graphs on a 5761-qubit QA for different graph topologies, 
indicating that more than 25\% of physical qubits remain unutilized, primarily due to dominant chains.

The severe underutilization of QA qubits, along with utilizing several physical qubits to represent a single program qubit, severely diminishes the capacity of QAs by up to 33x.
For instance, while current D-Wave QAs boast over 5,700 qubits, they can accommodate at most 177 program qubits with full connectivity.
The aim of this paper is to enable QAs to tackle larger problems by pruning dominant chains, while also enhancing the fidelity of the QAs.

\newpage
\section{Methodology}

\subsection{Hardware Platform}

For our evaluations, we utilize the D-Wave Advantage System (version 6.2), which features over 5,760 qubits and more than 40,000 couplers, accessed via the D-Wave Leap cloud service~\cite{D-Wave}.
We employ the default annealing time of 20 microseconds and adhere to the anneal schedule recommended for this device. 
Each problem is run for 4,000 trials to comply with the two seconds maximum job duration limit.

\subsection{Software Platform}

We utilize the \emph{minorminer} tool~\cite{minorminerGithub,cai2014practical} to find embeddings for arbitrary problem Hamiltonians on QA hardware. 
In our experiments, we set a timeout of 1,000 seconds, a maximum of 20 failed attempts for improvement, and conduct 20 trials. 
To program the D-Wave QAs, we employ the Ocean SDK~\cite{dwave_ocean_github}.

\subsection{Benchmarking}

Although current QAs feature over 5,700 qubits, their single-instruction operation model limits them to a few hundred program qubits with higher degrees, which is far below the number of variables required for real-world applications. 
Consequently, in this study, we employ synthetic benchmarks instead of real-world problems. 
In many real-world applications, graphs often exhibit a ``Power-Law'' distribution~\cite{agler2016microbial,clauset2016colorado,gamermann2019comprehensive,goh2002classification,house2015testing,mislove2007measurement,pastor2015epidemic}, 
and the \emph{Barabasi--Albert} (BA) algorithm~\cite{albert2005scale,barabasi1999emergence} is considered representative of these real-world graph structures~\cite{ayanzadeh2023frozenqubits,barabasi2000scale,gray2018super,kim2022sparsity,lusseau2003emergent,wang2019complex,zadorozhnyi2012structural,zbinden2020embedding}. 
The BA graphs are generated with a preferential attachment factor $m$, enabling us to vary the density of the graphs by adjusting $m$---with higher values of $m$ yielding denser graphs. 
We generate BA graphs with $m$ values ranging from $m=1$ (BA-1)  to $m=6$ (BA-6) to capture a broad spectrum of topologies, from sparse to nearly fully connected networks, thus effectively representing the dynamics of various real-world systems~\cite{clauset2016colorado}. 
Edge weights are assigned randomly following a standard normal distribution, which is a common approach in QA benchmarking~\cite{das2008colloquium,ayanzadeh2022equal,ayanzadeh2021multi}.

\subsection{Figure of merit}

In our evaluations, we use the \emph{Energy Residual} (\emph{ER}) to assess the fidelity of QA as 
\begin{equation}	
	\downarrow \textrm{Energy Residual (ER)} = \left| E_{min} - E_{global} \right|,
	\label{eq:ER}
\end{equation}	
where $E_{global}$ represents the global minimum of the benchmark problem, and $E_{min}$ corresponds to the best solution obtained by the QA.
Ideally, an ER value closer to zero is desirable as it indicates a solution that closely aligns with the ground state of the problem Hamiltonian.
We conducted intensive classical computations using state-of-the-art tools~\cite{ayanzadeh_ramin_2021_5142230} to determine the global minima of the benchmarks.

\newpage

\section{Skipper: Skipping Dominant Chains}
We introduce \emph{Skipper}, a software technique to enhance the capacity and fidelity of QAs through strategically skipping dominant qubit chains.

\subsection{Key Insights} \label{subsec:method_insights}

\subsubsection{Not All Program Qubits are Equal}

In digital QCs, the individual fabrication of physical qubits, such as superconducting ones, results in inevitable performance variations~\cite{tannu2019not}. 
Compilers, therefore, aim to prioritizing high-quality ones and limit the reliance on those of lower quality~\cite{tannu2019not,li2018tackling,noiseadaptive}.
However, in analog QCs, our observations reveal a significant variability at the level of program qubits.

\begin{figure}[h]
    \captionsetup[subfigure]{position=top} 
    \centering
    \subfloat[]{
        \includegraphics[width=0.49\columnwidth]{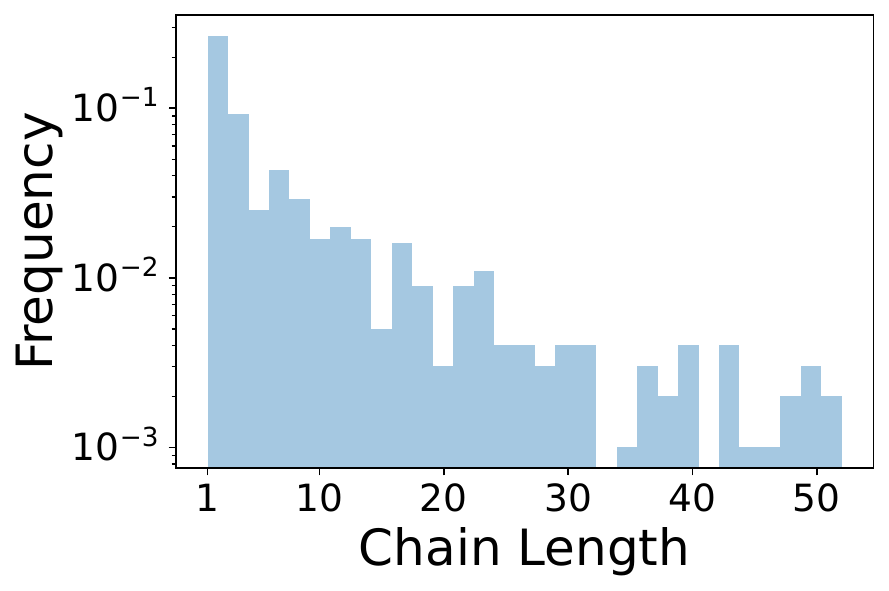}
    
    }
    \subfloat[]{
        \includegraphics[width=0.45\columnwidth]{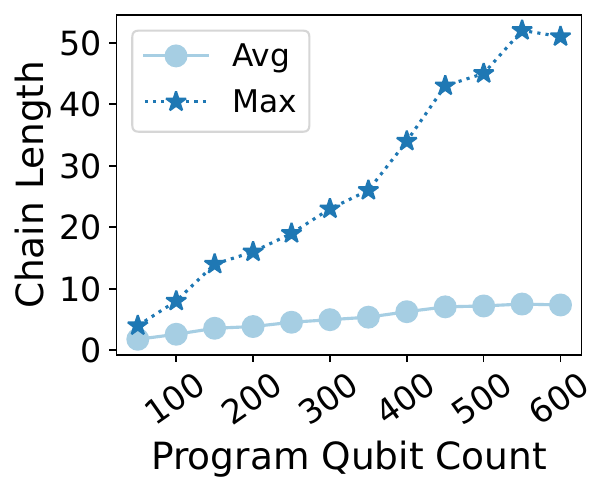}
    }
    \caption{  
(a) Histogram of chain lengths for a 600-node BA-3 graph (log-scale), indicating a ``Power-Law`` distribution of chain lengths.
 (b) Max and Avg chain lengths of BA-3 graphs, embedded on a 5761-qubit QA.
    }
    \label{fig:not_equal_qubits} 
\end{figure}

Figure~\ref{fig:not_equal_qubits}(a) shows the histogram of chain lengths (in log-scale) for the BA-3 graph type after embedding onto a 5761-qubit QA device, revealing a \emph{Power-Law} distribution with some notably longer \emph{dominant chains} and a majority of considerably shorter chains.
Figure~\ref{fig:not_equal_qubits}(b) presents the maximum and average chain lengths as the number of nodes in BA-3 graphs increases, notably magnifying the variability in chain lengths with the increase in problem size.

These intriguing observations extends beyond the BA-3 graph type, and we observe it in all benchmark graphs, including BA-1 to BA-6, spanning from sparse to nearly fully connected graphs. 
Additionally, we observe the nonuniformity of chain lengths in regular and fully connected graphs.


\subsubsection{QA Qubits are significantly Underutilized}

We observe that, on average, 25\% of physical qubits remain unused as they get trapped by chains.  
Additionally, we observe that the dominant chains significantly contribute to this qubit isolation.
This underutilization of QA qubits, along with utilizing several physical qubits to represent a single program qubit, severely diminishes the capacity of QAs by up to 33x.
For instance, the 2048-qubit and 5760-qubit QAs by D-Wave can accommodate a maximum of 64 and 177 fully connected program qubits, respectively.


\subsubsection{Diminishing Returns with Increased QA Trials}

QAs are noisy and prone to errors, leading to a systematic bias during the execution of quantum programs.
This bias causes deviations from the global optimum, reducing the reliability of QAs~\cite{albash2018adiabatic,mcgeoch2020theory,ayanzadeh2022equal,ayanzadeh2021multi}.
The bias arises from repeating the same quantum program across multiple iterations, exposing all trials to a similar noise profile~\cite{ayanzadeh2022equal}. 

\begin{figure}[h]    
    \centering
    \includegraphics[width=\columnwidth]{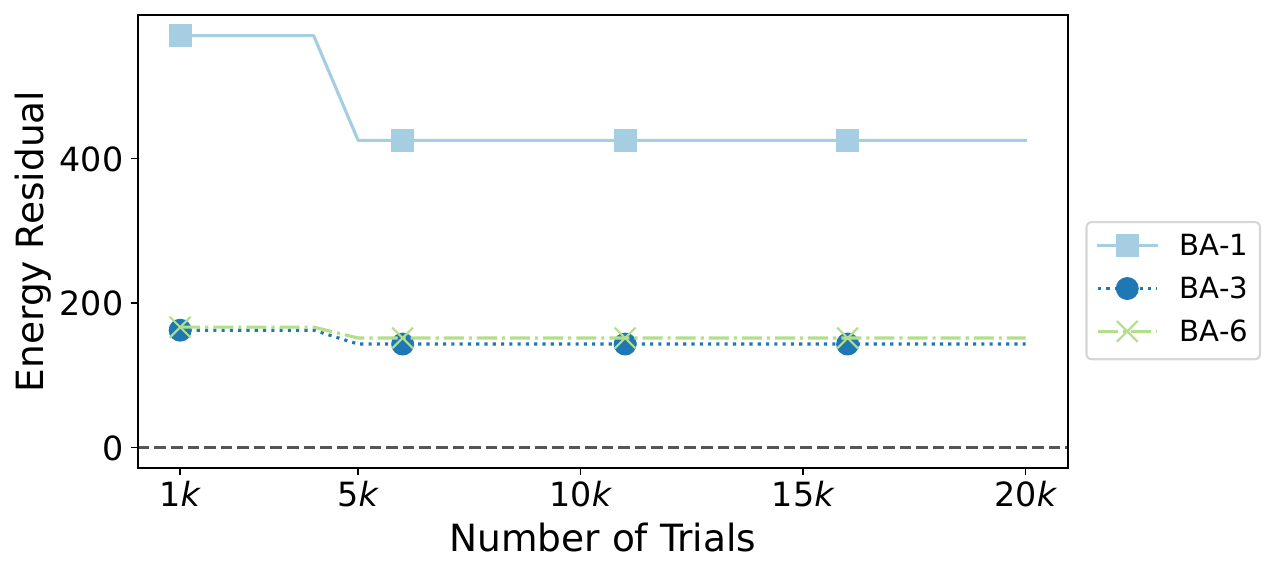}    
    \caption{        
        The Energy Residual (ER) in QAs tends to plateau with an increasing number of trials, and the global minimum often remains unreachable by QAs.
}    
    \label{fig:ER_plateau} 
\end{figure}

Figure~\ref{fig:ER_plateau} shows that when the number of trials in QA is increased, the output distribution reaches saturation. 
This indicates that the gap between the ideal solution and the QA does not reduce despite drawing more samples.
Moreover, due to the operation model of QAs as single-instruction computers, strategies commonly used in gate-based QCs~\cite{tannu2019ensemble,patel2020veritas} to address this bias are not applicable.


\begin{figure*}[t]    
    \centering    
    \includegraphics[width=1\textwidth]{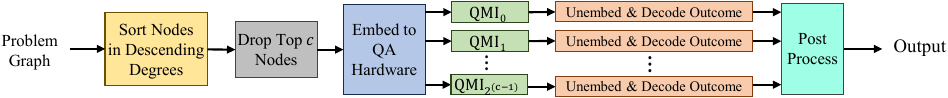}
    \caption{
    Overview of Skipper. 
    }
    \label{fig:m1_overview} 
\end{figure*}

\subsection{Overview of Skipper}

Figure~\ref{fig:m1_overview} shows the overview of Skipper. 
Skipper leverages insights into the distribution of chain lengths and the severe underutilization of qubits in QAs, employing a strategic approach to prune dominant chains and replace the corresponding program qubit with two potential measurement outcomes (+1 and -1). 
This process involves eliminating each chain, which partitions the search space into two sub-spaces. 
Skipper explores all sub-spaces, guaranteeing an exact recovery of the optimum solutions. 

Eliminating a dominant chain accomplishes two significant objectives: firstly, it frees up physical qubits previously used within pruned chains, and secondly, it eliminates the isolation of solitary qubits resulting from dominant chains.
As a result, Skipper enables the handling of larger problems by accommodating a significantly higher number of program qubits. 
Additionally, Skipper significantly enhances QA fidelity by substantially mitigating the impact of dominant chains, a primary factor in compromising QA reliability.
While Skipper utilizes more quantum resources due to the need to execute $2^c$ unique quantum programs for the removal of $c$ chains, it doesn't correspondingly enhance QAs (baseline) performance, as demonstrated in Fig.~\ref{fig:ER_plateau}.

\subsection{Chain Skips: How, Where, and When to Skip?}

\begin{figure}[h]
    \centering
    \includegraphics[width=\columnwidth]{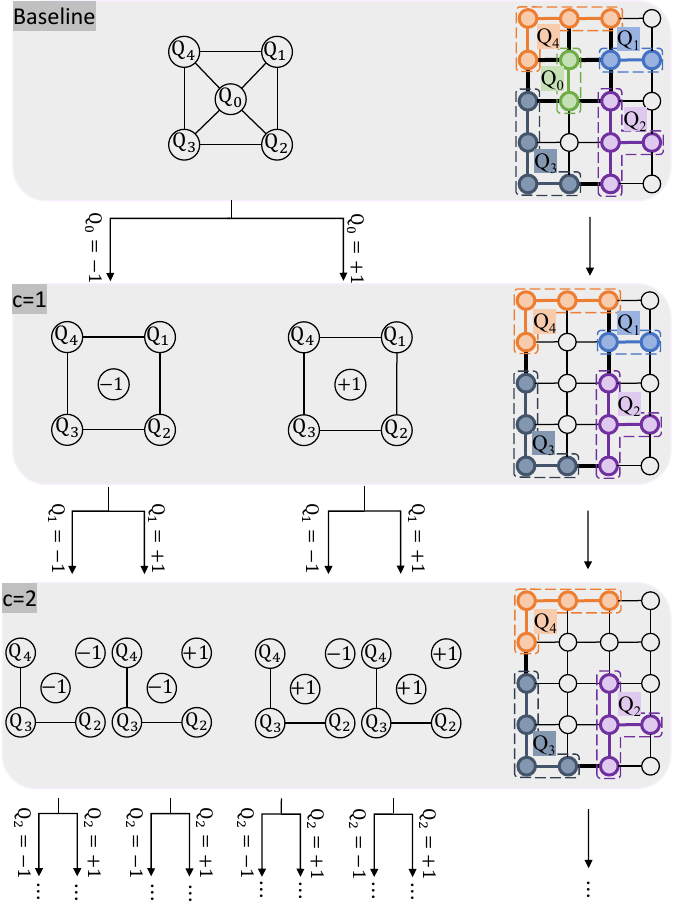}
    \caption{        
        By replacing $Q_0$ with +1 and -1 among five spin variables (baseline), two sub-problems each with four spin variables are obtained ($c=1$). 
        Fixing $Q_1$ in these two sub-problems with +1 and -1 results in four sub-problems with three spin variables ($c=2$).
        The same embedding is utilized for all sub-problems at each level of the binary tree. 
}           
    \label{fig:chain_cut_binary_tree} 
\end{figure}

Figure~\ref{fig:chain_cut_binary_tree} illustrates the elimination of two chains from a problem with five variables, creating two and four \emph{independent} sub-problems, respectively. 
To skip a chain, the program qubit is replaced with +1 and -1 (two measurement outcomes), removing the node and its connected edges from the problem graph.
Unlike digital QCs, where removing one program qubit results in reducing the physical qubit utilization by one, in QAs, removing one program qubit liberates all the physical qubits involved in its corresponding chain.

\begin{figure}[ht]
    \centering
    \includegraphics[width=\columnwidth]{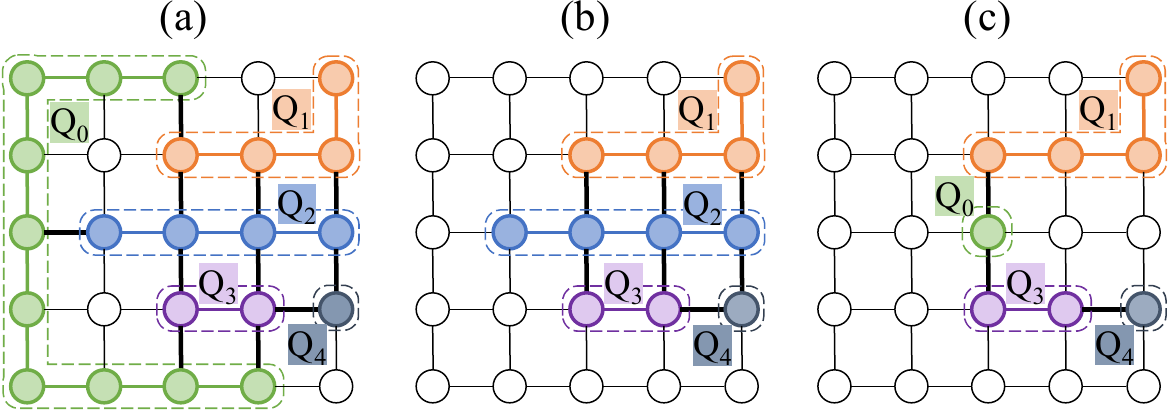}
    \caption{  
(a) Embedding of five program qubits on a grid. 
(b) Freeing ten qubits by pruning the dominant chain $Q_0$. 
(c) Fourteen qubits freed by pruning $Q_2$. 
    }
    \label{fig:where_to_cut} 
\end{figure}

Identifying dominant chains to trim in Skipper is nontrivial.
In digital QCs, high-degree program qubits necessitate more CNOT gates, enabling direct identification prior to circuit compilation. 
However, in QAs, it is not feasible to directly recognize program qubits linked to longer chains, thus requiring embedding techniques to identify them.
Furthermore, it is not always optimal to prune the dominant chain. 
In Fig.~\ref{fig:where_to_cut}(a), the dominant chain is $Q_0$ and consists of ten physical qubits. 
Pruning $Q_0$ (Fig.~\ref{fig:where_to_cut}(b)), liberates all ten physical qubits, leaving the other chains intact. 
However, as shown in Fig.~\ref{fig:where_to_cut}(c), removing $Q_2$ and re-embedding the problem not only releases the five physical qubits associated with $Q_2$ but also effectively reduces $Q_0$ to a singleton chain, totaling fourteen qubits released.

Skipper adopts a greedy approach to prune $c$ chains by sorting program qubits based on their degree and removing the top $c$ qubits simultaneously. The removal of a single program qubit can have a substantial impact on other chains, as shown in Fig.~\ref{fig:where_to_cut}(c). 
In the context of irregular graphs that often follow the Power-Law distribution in real-world applications, this greedy approach exhibits a desirable, near-optimal behavior for $c \geq 5$ chain cuts.

\subsection{Skip Count: A Cost-Performance Tradeoff}

Skipper permits users to trim up to eleven chains. 
Each chain skipped bifurcates the search space; therefore, trimming up to eleven chains can lead to a maximum of 2048 sub-problems. 
Skipper executes all corresponding QMIs to ensure exact solution recovery.
However, the nontrivial embedding process and the need to execute up to 2048 embeddings can create a bottleneck for Skipper. 
Fortunately, the identical structure of all sub-problems at the $c$-th level in the binary tree enables sharing the same embedding across them (Fig.~\ref{fig:chain_cut_binary_tree}).

\subsection{Unembedding: Remediating Broken Chains}

The sub-problem resulting from cutting chains consists of $n-c$ program qubits.
However, after embedding, the problem executed on the QA hardware encompasses $N$ physical qubits, where $c \ll n \ll N$.
As a result, the QA produces outcomes as $N$-bit strings, with each program qubit collectively represented by multiple bits in a chain.
Therefore, Skipper \emph{unembeds} these outcome samples, converting them back into the space of program qubits.

\begin{figure}[h]
    \centering
    \includegraphics[width=\columnwidth]{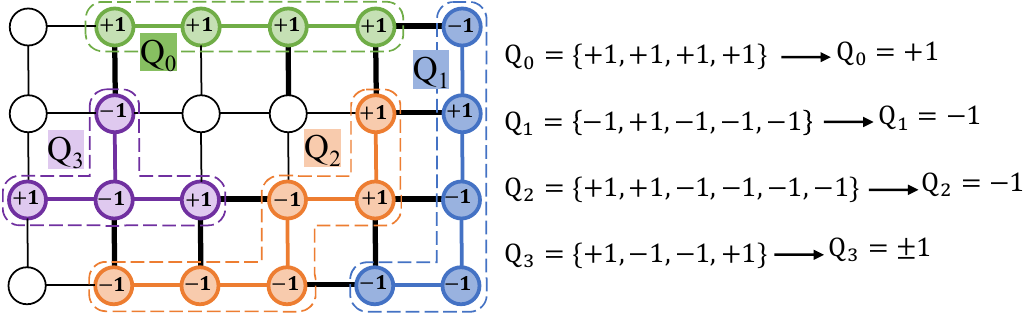}
    \caption{  
Unembedding examples
    }
    \label{fig:unembedding_example} 
\end{figure}

Ideally, all physical qubits within a chain should have identical values in a given QA sample. 
The value of the associated program qubit is then determined by observing any one of the physical qubits within it (e.g., program qubit $Q_0$ in Fig.~\ref{fig:unembedding_example}).
However, QAs are inherently open systems, as interactions with the environment are unavoidable in QCs, and the annealing process tends to be diabatic since truly adiabatic processes are often unfeasible~\cite{ayanzadeh2021multi}.
As a result, qubits within a chain can take different values, an issue known as \emph{broken chains}~\cite{grant2022benchmarking,pelofske2020advanced,king2014algorithm,barbosa2021optimizing}.

To remediate broken chains, Skipper employ the \emph{majority voting} approach. 
For instance, in Fig.~\ref{fig:unembedding_example}, although $Q_1$ exhibits a broken chain with varying qubit values, the unembedding process assigns a value of -1, reflecting the majority of -1 values within the chain (4 versus 1).
However, not all chains have an odd length, and forcing the embedding to produce odd chain lengths is nontrivial.
Unembedding even length chains with mostly identical qubit values (e.g., $Q_2$ in Fig.~\ref{fig:unembedding_example}) is not challenging, as majority voting can effectively determine the value of the program qubit.
However, as demonstrated by $Q_3$ in Fig.~\ref{fig:unembedding_example}, a chain of even length can contain an equal number of -1 and +1 values, referred to as \emph{balanced chains}, a condition where majority voting fails.
Skipper manages balanced chains by counting them and implementing distinct strategies based on their quantity. 
For problems with fewer than ten balanced chains, Skipper discards their qubit values and uses a brute-force approach (with up to 1024 possible configurations), selecting the configuration that yields the lowest energy value. 
If the number of balanced chains exceeds ten, Skipper randomly assigns values to the corresponding program qubits. 
When a broken chain occurs, Skipper can optionally apply Single-Qubit Correction (SQC)~\cite{ayanzadeh2021multi,ayanzadeh2022equal} postprocessing to maintain a feasible solution for the original problem.

\subsection{Decoding Sub-Problem Results}

After unembedding, each sample will encompass $n-c$ bits, while the original problem includes $n$ variables. 
The decoding process reintroduces the values of the $c$ pruned program qubits,
which were fixed during the sub-problem formulation by assigning fixed values to these variables. 

\subsection{Postprocessing}

Theoretically, QAs sample from a Boltzmann distribution, exponentially favoring lower energy values, and thus should locate the global optimum in few attempts. 
However, like other QCs, QAs are vulnerable to noise and various error sources that degrade their fidelity. 
To enhance the reliability of QAs, we can optionally apply postprocessing heuristics to the resulting samples~\cite{ayanzadeh2021multi}.

\subsection{Deriving the Final Output}
In Skipper, all sub-problems are executed independently, each one corresponding to a separate sub-space of the primary problem. 
Consequently, in Skipper, the sample with the lowest energy or objective value is deemed as the ultimate output, with the originating sub-space of this global optimum being of no consequence.

\subsection{Overhead of Skipper}

Let ${c}$ represent the number of skipped chains, $e$ denote the edges in the problem graph, $r$ symbolize the number of trials on the QA, while $n$ and $N$ correspond to the number of program and physical qubits, respectively.

\vspace{0.05in}
\noindent \textbf{Quantum overhead:}
Skipper allows for up to eleven chain cuts, necessitating the execution of at most 2048 distinct quantum executables, each running independently.

\vspace{0.05in}
\noindent \textbf{Classical overhead:}
We separate the embedding overhead of Skipper from all other classical pre/post-processing modules, as the embedding is the primary factor influencing the end-to-end runtime of the proposed techniques in this paper (refer to section~\ref{sec:workflow}). 
Given the fact that $c \ll n \ll N \ll r$, Skipper demonstrates a classical time complexity of $O\left( 2^c \left(rN + c \right) \right)$. 
This is representative of the unembedding and decoding processes for outcome samples of sub-problems.

\vspace{0.05in}
\noindent \textbf{Embedding overhead:}
In Skipper, all sub-problems of the binary tree at the $c$-th level share a single embedding, leading to $O(1)$ embedding complexity. 
Note that we assume all sub-problems are executed on the same QA hardware or that all devices have the same working graph topology, allowing them to share the embedding.

\vspace{0.05in}
\noindent \textbf{Memory utilization:}
The memory utilization in Skipper scales according to $O(rN2^c)$.

\newpage
\section{Skipper Evaluation Results}
We evaluate Skipper using Barabasi--Albert (BA) graphs~\cite{barabasi1999emergence} 
with different preferential attachment factor values: $m=1$ (BA-1) to $m=6$ (BA-6). 

\subsection{Solving Larger Problems}

\subsubsection{Impact on Chain Length}

\begin{figure}[h]
    \captionsetup[subfigure]{position=top} 
    \centering
    \subfloat[]{
		\includegraphics[width=0.49\columnwidth]{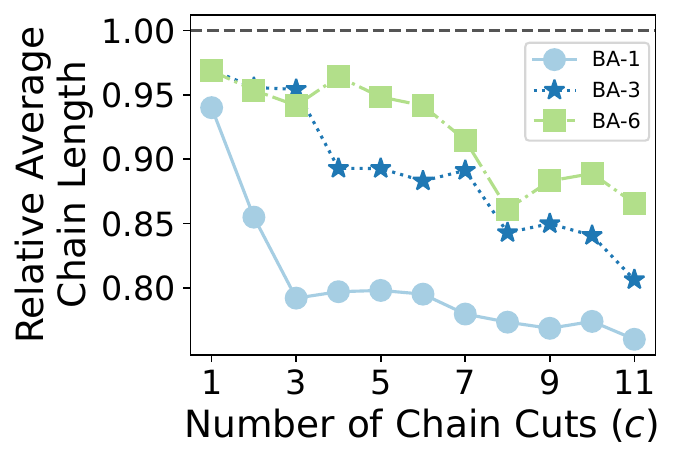}
	}
    \subfloat[]{
		\includegraphics[width=0.48\columnwidth]{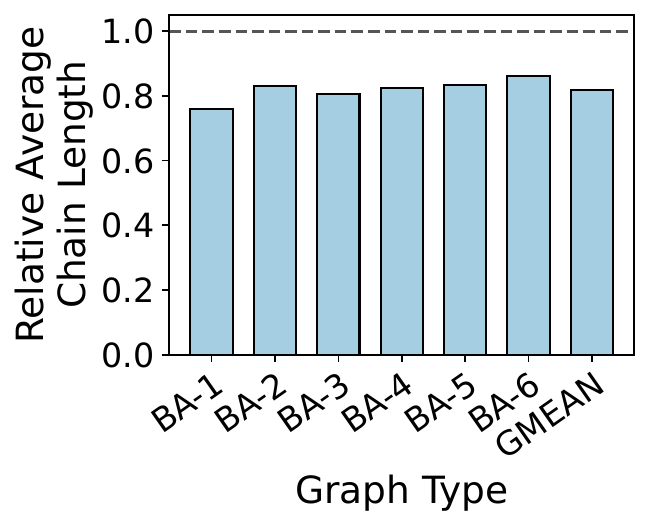}
	}
    \caption{        
        Relative Avg. chain length in Skipper compared to baseline (lower is better). 
        (a) Relative Avg. chain length for different graphs as cut size ($c$) increase. 
        (b) Overall relative mean chain lengths for up to 11 chain cuts. 
}    
    \label{fig:future_relative_avg_chaining_cost} 
\end{figure}

Figure~\ref{fig:future_relative_avg_chaining_cost}(a) illustrates that increasing the number of chain cuts ($c$) in Skipper leads to a reduction in the average chain length of the embeddings.
Figure~\ref{fig:future_relative_avg_chaining_cost}(b) demonstrates that Skipper decreases the mean chain length by up to 1.32x (with an average of 1.22x) when cutting up to eleven chains.

\begin{figure}[t]
    \captionsetup[subfigure]{position=top} 
    \centering
    \subfloat[]{
		\includegraphics[width=0.49\columnwidth]{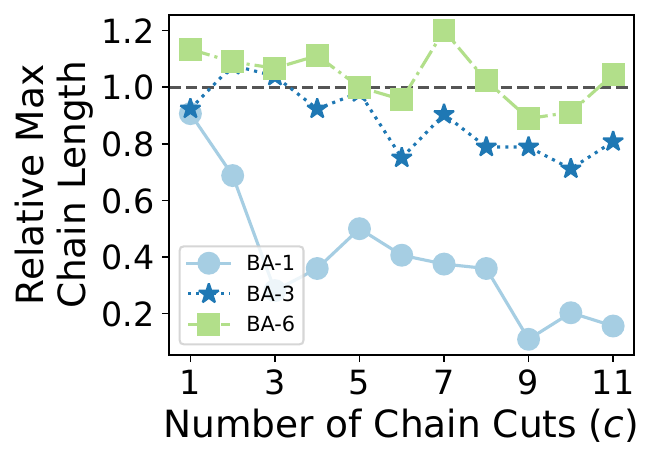}
	}
    \subfloat[]{
		\includegraphics[width=0.49\columnwidth]{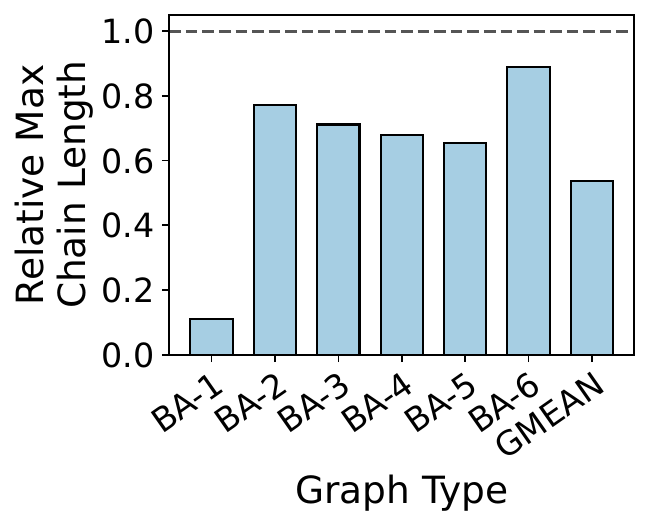}
	}
    \caption{        
        Relative Max chain length in Skipper compared to the baseline (lower is better). 
        (a) Relative Max chain length for different graphs as $c$ increases.
        (b) Overall relative max chain lengths for up to 11 chain cuts.
    }    
    \label{fig:future_relative_max_chaining_cost}
\end{figure}

Figure~\ref{fig:future_relative_max_chaining_cost}(a) shows that the maximum chain length of the embeddings decreases as $c$ in Skipper increases.
Figure~\ref{fig:future_relative_max_chaining_cost}(b) shows that cutting up to $c=11$ chains in Skipper reduces the maximum chain length by up to 9.14x (average 1.86x).
Our observations indicate that long chains are the primary contributing factor to the underutilization of physical qubits.

\subsubsection{Impact on Qubit Utilization}

\begin{figure}[h]
    \captionsetup[subfigure]{position=top} 
    \centering
    \subfloat[]{
		\includegraphics[width=0.50\columnwidth]{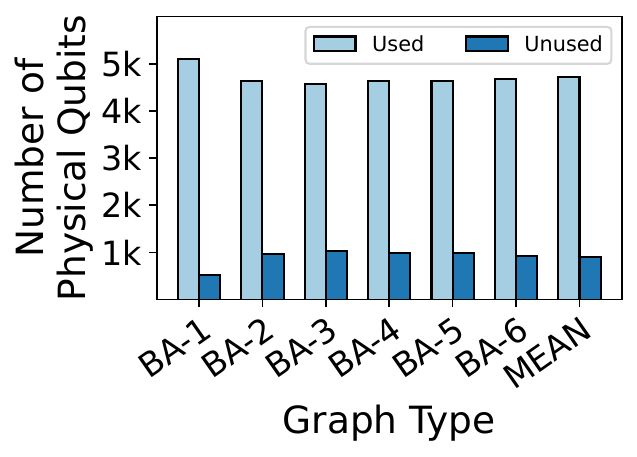}
	}
    \subfloat[]{
		\includegraphics[width=0.48\columnwidth]{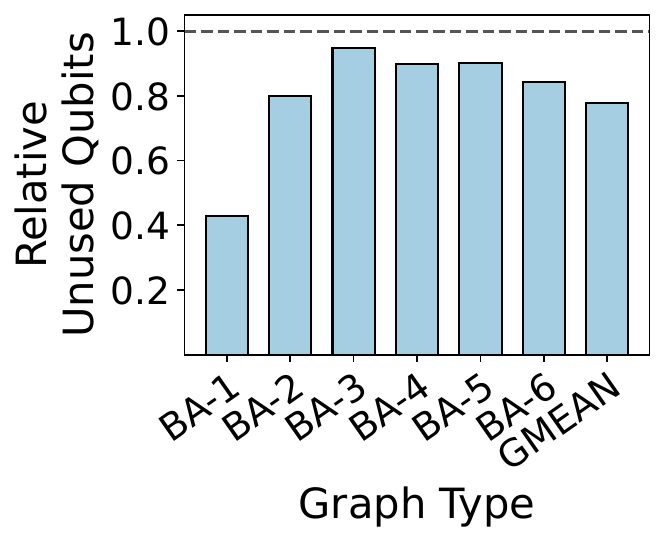}
		\label{subfig:future_relative_unused_qubits}
	}
    \caption{         
        (a) Utilization of Physical Qubits in Skipper across Different Graph Types.
        (b) Relative Number of Unused Physical Qubits in Skipper for up to 11 Chain Cuts, Compared to the Baseline.
Lower is better. 
    }    
    \label{fig:future_qubit_utilization} 
\end{figure}

Figure~\ref{fig:future_qubit_utilization}(a) displays the average and maximum number of physical qubits when up to eleven chains are pruned. 
In Fig.~\ref{fig:future_qubit_utilization}(b), Skipper reduces underutilization of QA qubits by up to 57\% (average 22.14\%) with up to eleven trimmed chains. 

\subsubsection{Impact on Capacity of QAs}

\begin{figure}[]
    \captionsetup[subfigure]{position=top} 
    \centering
    \subfloat[]{
		\includegraphics[width=0.48\columnwidth]{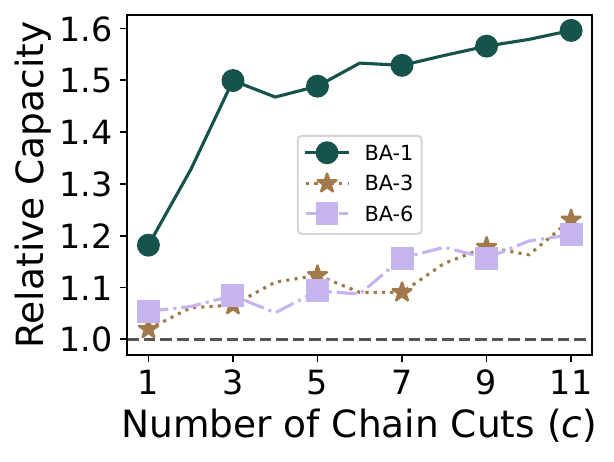}
	}
    \subfloat[]{
		\includegraphics[width=0.49\columnwidth]{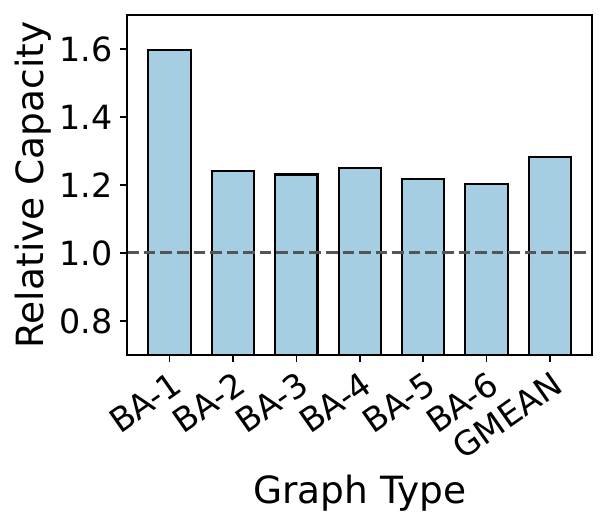}
		\label{subfig:future_relative_capacity}
	}
    \caption{        
        Relative QA capacity in Skipper compared to baseline. 
         (a) Relative capacity for different graphs as cuts increase. 
         (b) Overall relative capacity for up to 11 chain cuts.
         Higher is better.
    }    
    \label{fig:future_relative_capacity} 
\end{figure}

The QA capacity to accommodate specific graph types, from BA-1 to BA-6, is determined by the largest number of program qubits of each type that can be embedded on the QA.
Figure~\ref{fig:future_relative_capacity}(a) shows that QA capacity in Skipper improves with increasing $c$ across various graph topologies.

Figure~\ref{fig:future_relative_capacity}(b) demonstrates that Skipper enables the embedding of larger problems onto current QAs, with an increase of up to 59.61\% (average 28.26\%). 
It is important to note that this growth in the number of program qubits necessitates a substantial increase in the number of physical qubits, as one program qubit is represented by multiple physical qubits. 

\begin{tcolorbox}[colback=blue!12]
	Skipper's performance remains consistent regardless of the increasing density of problem graphs (from BA2 to BA6). 
\end{tcolorbox}


\newpage 
\subsection{Boosting QA Reliability} \label{subsec:m1_reliability}
In addition to enhancing QA capacity, Skipper can be employed to improve the reliability of QAs.

\subsubsection{Impact on Embedding Quality }

\begin{figure}[ht]
    \captionsetup[subfigure]{position=top} 
    \centering
    \subfloat[]{
		\includegraphics[width=0.48\columnwidth]{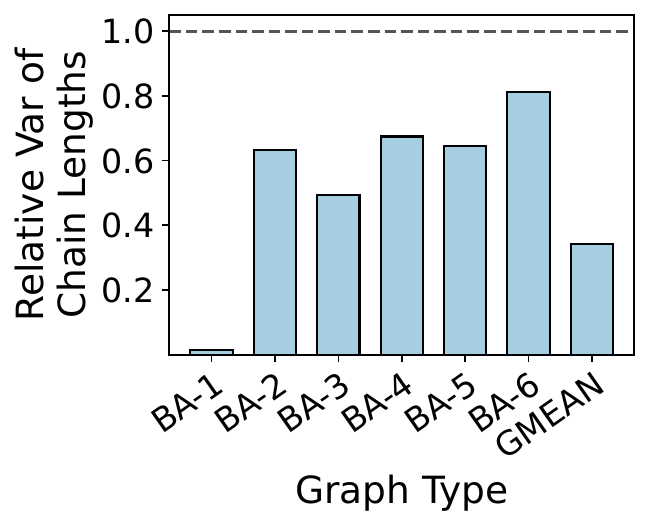}
	}
    \subfloat[]{
		\includegraphics[width=0.48\columnwidth]{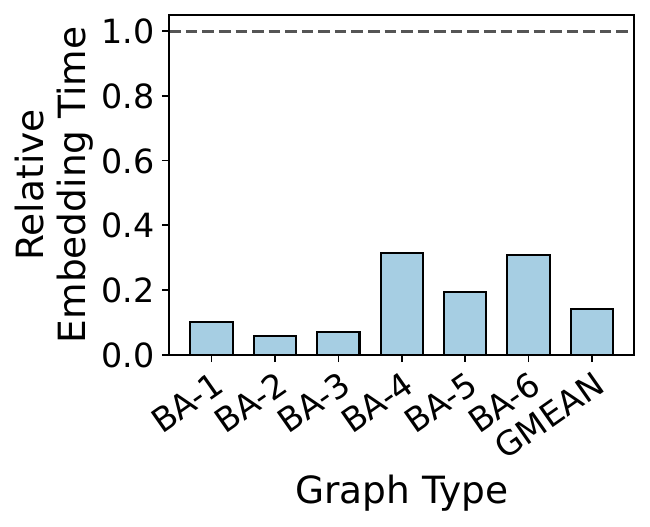}
		\label{subfig:future_relative_capacity}
	}
    \caption{ 
(a) Relative variance of chain lengths and (b) relative embedding time in Skipper compared to the baseline when trimming up to eleven chains. 
Lower is better. 
}
    \label{fig:future_embedding_performance} 
\end{figure}

QAs do not incorporate circuits, thus precluding the use of the Probability of Successful Trials metric commonly employed to assess compilation quality in digital QCs~\cite{ayanzadeh2023frozenqubits, alam2020circuit, nishio, tannu2022hammer}.
Prior studies suggest that embeddings with similar chain lengths can produce better solutions~\cite{boothby2016fast,venturelli2015quantum,rieffel2015case,choi2008minor}.
Figure~\ref{fig:future_embedding_performance}(a) demonstrates that trimming up to eleven chains in Skipper reduces the average variance in chain lengths by 2.93x (up to 70.19x).

\subsubsection{Impact on Embedding Time}

Figure~\ref{fig:future_embedding_performance}(b) demonstrates that pruning up to eleven chains in Skipper leads to a significant reduction in embedding time, with a maximum improvement of 17.13x (average improvement of 7.12x).

\subsubsection{Impact on Fidelity}

\begin{figure}[h]
    \captionsetup[subfigure]{position=top} 
    \centering
    \subfloat[]{
		\includegraphics[width=0.48\columnwidth]{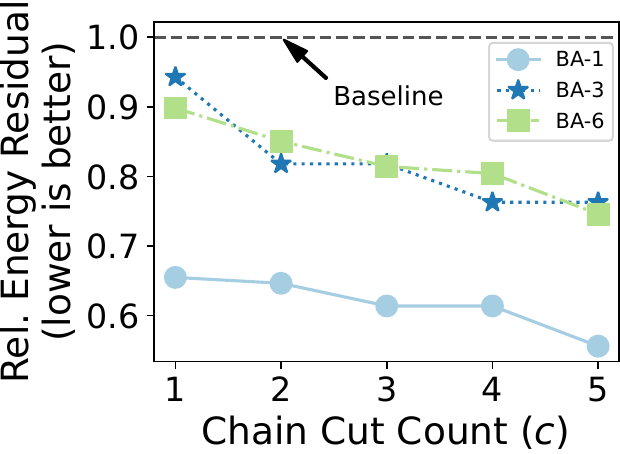}		
	}
    \subfloat[]{
		\includegraphics[width=0.48\columnwidth]{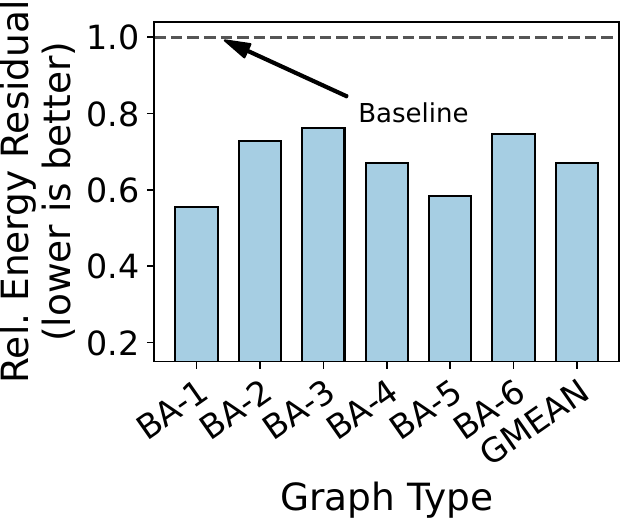}
	}
    \caption{  
        Relative Energy Residual (ER) in Skipper compared to baseline (lower is better). 
        (a) Relative ER as $c$ increase. 
        (b) Overall relative ER for up to five chain cuts.
    }
    \label{fig:current_fidelity} 
\end{figure}

Figure~\ref{fig:current_fidelity}(a) shows that as Skipper skips more chains, the Energy Residual (ER) decreases, indicating a progressive approach towards the global optimum.
Additionally, Fig.~\ref{fig:current_fidelity}(b) demonstrates a significant reduction in ER by up to 44.4\% (average 33.08\%), when up to five chains are cut using Skipper, compared to the baseline.



\newpage
\section{Skipper-G: A Marcovian Approach}

We propose \emph{Skipper-G}, a greedy scheme that reduces the quantum cost of Skipper by skipping the examination of sub-problems unlikely to contain the global optimum. 
However, this strategy entails a trade-off: Skipper-G achieves marginally lower fidelity gains compared to Skipper and is ineffective for enhancing QA capacity.

\subsection{Insight: Not ALL Sub-Spaces Include Global Optimum}

\begin{figure}[b]
    \centering
    \includegraphics[width=0.9\columnwidth]{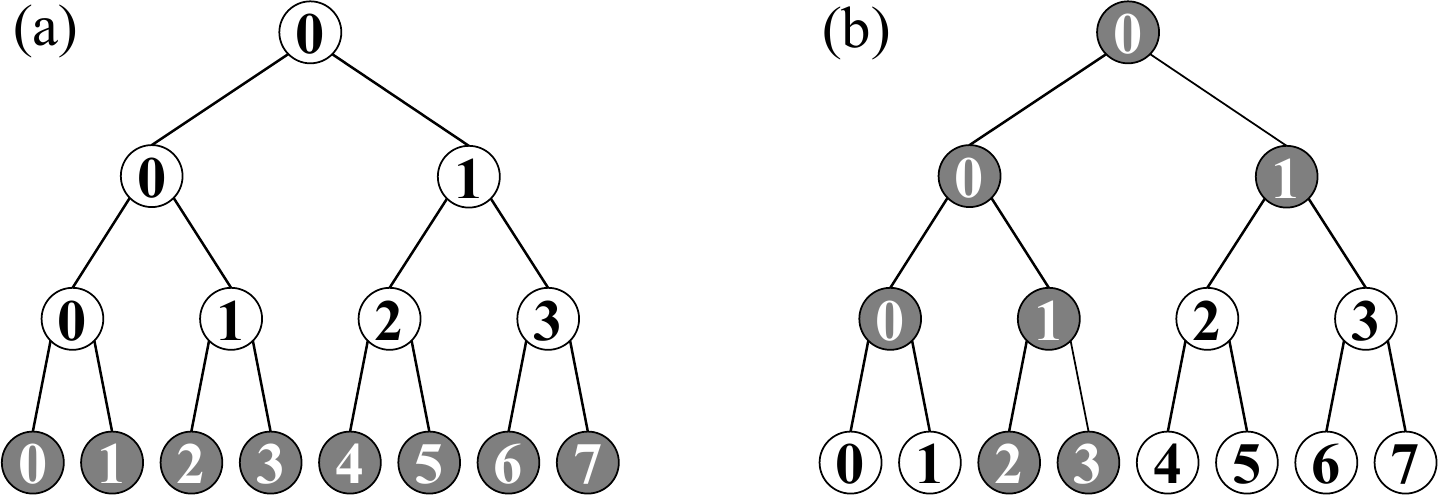}
    \caption{
        (a) Skipper utilizes a Breadth-First Search (BFS) strategy, examining all leaf nodes (intermediate nodes are not examined).
        (b) Skipper-G adopts a Depth-First Search (DFS) strategy, examining only two nodes at each level of the binary tree (including intermediate nodes).
        Example: With $c=11$ chain cuts, Skipper and Skipper-G execute at most 2048 and 23 QMIs, respectively.
}       
    \label{fig:BFS_DFS} 
\end{figure}

Skipper employs a Breadth-First Search (BFS) strategy to examine sub-problems, as depicted in Fig.~\ref{fig:BFS_DFS}(a). 
Trimming each chain bifurcates the search space, with skipping $c$ chains resulting in a binary tree of depth $c$. 
To ensure successful recovery, Skipper evaluates all leaf nodes, running a separate QMI for each sub-space at the tree's last level. 
Notably, Skipper does not examine intermediate nodes (or sub-spaces) since all chains are trimmed simultaneously. 

Users define the number of chain cuts in Skipper, with the option to skip up to eleven chains based on their budgetary constraints. 
For instance, if a user opts for the maximum allowable eleven cuts, Skipper must run 1024 QMIs when all linear coefficients are zero~\cite{ayanzadeh2023frozenqubits}, and up to 2048 QMIs otherwise.
Nonetheless, these sub-problems are independent, allowing for parallel execution by Skipper. 
Notably, Skipper's overall runtime remains comparable to the baseline, attributed to the significantly reduced embedding time, as detailed in Section~\ref{subsec:m1_reliability}.
However, the quantum costs incurred on QCs are substantially higher than those on classical platforms, which may present affordability issues for some users.

Not every sub-space contains the global optimum. 
Leveraging this insight, we introduce \emph{Skipper-G} (\emph{greedy Skipper}), which reduces the quantum cost of Skipper by adopting a Depth-First Search (DFS) strategy (Fig.~\ref{fig:BFS_DFS}(b)), to bypass sub-spaces unlikely to include the global optimum.
When pruning the maximum of eleven chains, Skipper-G executes 23 QMIs, in contrast to Skipper's potential 2048 QMIs.

\subsection{How Skipper-G Work?}

\begin{figure}[t]
    \centering
    \includegraphics[width=1\columnwidth]{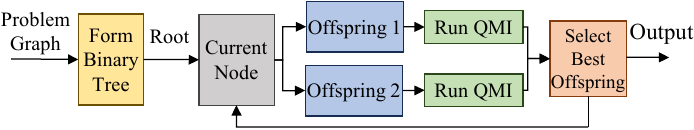}
    \caption{
Overview  of Skipper-G. 
}       
    \label{fig:m2_overview}
\end{figure}

Figure \ref{fig:m2_overview} illustrates the overview of the Skipper-G scheme. 
In Skipper-G, similar to Skipper, users can determine the number of chain cuts, with the possibility of skipping up to eleven chains, depending on their budget constraints.
However, unlike Skipper where all chains are cut simultaneously, Skipper-G employs an iterative approach, cutting one chain in each iteration.
As illustrated in Fig.~\ref{fig:BFS_DFS}(b), Skipper-G initiates by setting the root node (i.e., the baseline with no chains cut) as the current node and executing the corresponding quantum program.
For each chain cut, Skipper-G performs the following steps:
\begin{enumerate}
    \item 
    In the current node (problem), the dominant chain is trimmed by setting its corresponding program qubit to either +1 or -1, resulting in two child nodes. 
    If the current node at level $c$ has the index $x$, then its left and right children at level $c+1$ will have indices $2x$ and $2x+1$, respectively (e.g., node $x=1$ at the third level leads to nodes 2 and 3 in Fig.~\ref{fig:BFS_DFS}(b)).
    
    \item 
    The quantum programs corresponding to the children are executed on the QA.
    
    \item 
    The best offspring is set as the current node.
\end{enumerate}


\subsection{Branch and Bound Criteria}


When evaluating a node in Skipper-G, a quantum program is executed on a QA device for multiple trials. 
Each trial produces an outcome with an associated objective value. 
The assessment of node quality in Skipper-G is based on the following feature (lower is better):
\begin{equation}
    \downarrow f(Z) = \left| \frac{1}{\text{E}_{\text{min}} \times \text{EV}} \right|,
    \label{eq:m2_f}
\end{equation}
where $Z$ denotes the set of obtained samples, and $\text{E}_{\text{min}}$ and $\text{EV}$ represent the minimum and the expected value of the energy values in $Z$, respectively. 
The lower the value of $f$, the greater the likelihood that a child includes the global optimum in its corresponding subspace during the traversal of the associated binary tree. 
This feature balances the best sample with the overall quality of all samples, reducing the likelihood of getting trapped in local optima.

\subsection{Overhead of Skipper-G}

Skipper-G is capable of trimming up to eleven chains, which necessitates a maximum of 23 distinct QMI executions.
Although Skipper-G examines two nodes at each level of the binary tree, these nodes, due to their identical structures, can utilize a single embedding. 
Consequently, Skipper-G requires $c$ embeddings for $c$ chain cuts. 
Nonetheless, since these embeddings can be executed in parallel, the embedding time for Skipper-G remains similar to that of the baseline, as the root node's embedding is expected to be more time-consuming than the embeddings of the smaller intermediate nodes.

\subsection{Evaluation Results}

\begin{figure}[h]
    \captionsetup[subfigure]{position=top} 
    \centering
    \subfloat[]{
        \includegraphics[width=0.48\columnwidth]{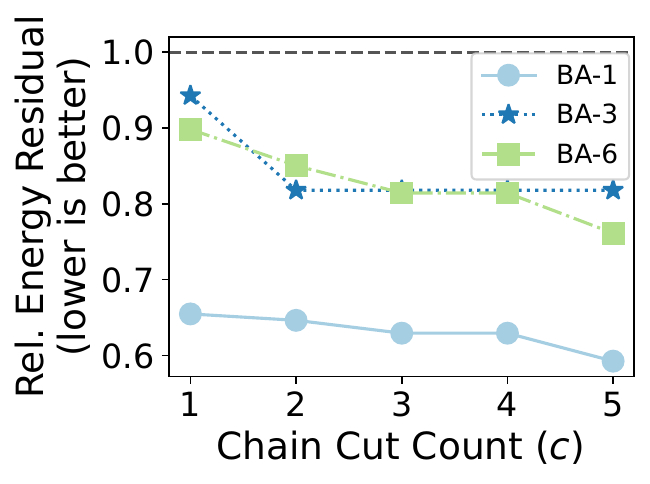}
    }
    \subfloat[]{
        \includegraphics[width=0.48\columnwidth]{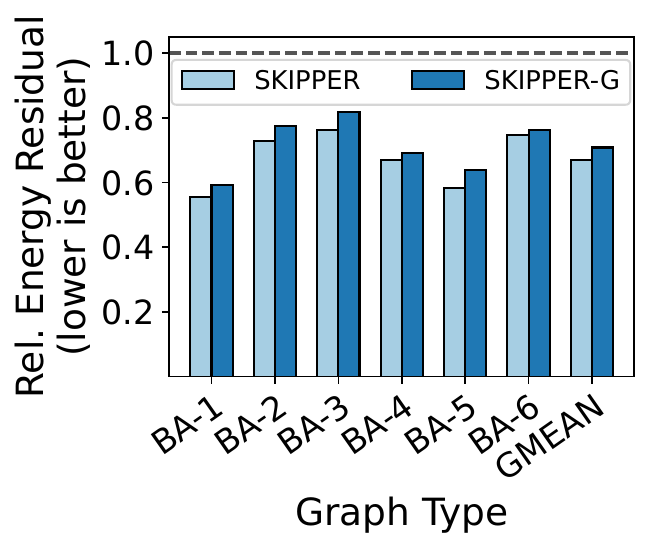}
    }
    \caption{  
        Relative ER for five chain cuts compared to the baseline (lower is better). 
        (a) Skipper-G: Relative ER with increasing $C$. 
        (b) Overall Relative ER (lower is better) for Skipper vs. Skipper-G.
    }
    \label{fig:DFS_fidelity} 
\end{figure}

Figure~\ref{fig:DFS_fidelity}(a) illustrates the ER for various chain cut counts in Skipper-G, showing that progressively trimming more dominant chains leads to a decrease in the ER, approaching the global minimum. 
Additionally, Fig~\ref{fig:DFS_fidelity}(b) reveals that pruning up to five chains in Skipper-G can reduce the gap between the global optimum and the best QA sample by as much as 40.75\% (Avg. 29.19\%) compared to the baseline. 

Skipper marginally outperforms Skipper-G, achieving a 3.89\% greater reduction in ER, albeit at the expense of significantly higher quantum resource utilization. 
Skipper-G includes the baseline at the root of the binary tree, ensuring it performs no worse than the baseline.

\section{Skipper and Skipper-G in Classical Realm}

Unfortunately, neither Skipper nor Skipper-G can be utilized to enhance the fidelity of optimization techniques used in classical realm. 
In the classical domain, the hardness of optimization problems depends on the number of variables and the graph topology. 
For instance, while planar graphs are tractable~\cite{dei2006exact,hadlock1975finding} in classical realm, neither regular nor power-law graphs become planar simply by eliminating a few nodes.
Additionally, eliminating ten nodes from a 1000-node graph results in sub-graphs with 990 nodes, which typically remain intractable in the classical realm.

Similarly, Skipper is not suitable for tackling larger problems in the classical realm. 
Its primary goal is to address the sparse connectivity of qubits, a key factor limiting the capacity of QAs. 
However, the full connectivity of classical bits does not present a similar limitation.

\newpage
\section{Workflow Analysis} \label{sec:workflow}

The runtime of quantum programs is mainly determined by queuing delays, execution modes through cloud services (which vary across providers), and embedding time, rather than the execution time on quantum hardware (microseconds to milliseconds). 
To offer a holistic examination of the runtime between the proposed techniques and the baseline, we employ the following analytical model:
\begin{equation}
	T = T_{\text{emb}} + N_{\text{QMI}} \left({ T_{\text{queue}} + T_{\text{QMI}} + T_{\text{net}} }\right) + T_{\text{classical}},
	\label{eq:overall_runtime}
\end{equation}
where $T_{\text{emb}}$ is the embedding time, $N_{\text{QMI}}$ is the number of quantum executables, $T_{\text{queue}}$ is the job wait time, $T_{\text{QMI}}$ is the QMI execution time, $T_{\text{net}}$ is the network delay, and $T_{\text{classical}}$ is the classical pre/post-processing time.

For $r$ trials, $T_{QMI} = t_p + \Delta + r \times t_s$, where $t_p$ is the raw signal preparation time, $\Delta$ is the 10ms QA initialization time, and $t_s$ is the single annealing/readout time. 
Given that D-Wave limits $T_{QMI}$ to two seconds, we assume $T_{QMI} = 2$ in all cases. 
We assume one second for $T_{\text{net}}$ for each job.

We assume a baseline embedding time of 30 minutes, decreasing proportionally with skipped chains (as discussed in section~\ref{subsec:m1_reliability}). 
For example, pruning ten chains reduces the embedding time to three minutes. 
All embeddings can be computed in parallel, making $T_{\text{emb}}$ in Skipper-G the maximum time for individual embeddings. 
Additionally, we allocate one second each for pre- and post-processing.

\begin{figure}[h]    
    \centering    
	\includegraphics[width=\columnwidth]{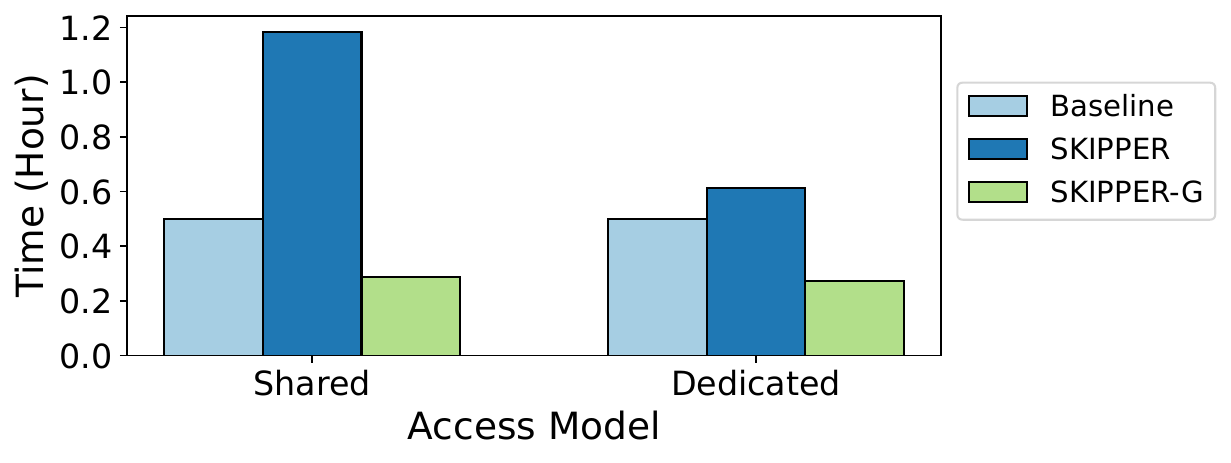}
    \caption{  
Overall Runtime comparison. 
    }
    \label{fig:latency} 
\end{figure}

We examine two access scenarios: \emph{shared} and \emph{dedicated}, with one and zero-second queuing times, respectively. 
Figure~\ref{fig:latency} compares the end-to-end runtime of the baseline and our proposed techniques with $c=11$, resulting in up to 1024 QA runs. 
Skipper shows significantly greater advantages over others in the dedicated access mode.

\ignore{
Table~\ref{tbl:methods} offers a comprehensive overview of the proposed techniques. 

\begin{table}[htbp]
    \centering
    \caption{
        Skipper family characteristics relative to the baseline. 
        }
    \centering
    \label{tbl:methods} 
    \begin{tabularx}{\columnwidth}{|l|l|l|}
      \cline{2-3}
      \multicolumn{1}{c|}{} & \textbf{Skipper} & \textbf{Skipper-G} \\
      \cline{2-3}      
      \cline{1-1}      
      \textbf{Quantum Cost} & Exponential & Linear \\
      \cline{1-3}
      \textbf{Embedding Cost} & No Cost& Linear \\
      \cline{1-3}
      \textbf{Avg Fidelity Improvement} & 33.08\% & 29.19\% \\
      \cline{1-3}
      \textbf{Max Fidelity Improvement} & 44.4\% & 40.75\% \\
      \cline{1-3}
      \textbf{Avg Capacity Improvement} & 28.26\% & NA \\
      \cline{1-3}
      \textbf{Max Capacity Improvement} & 59.61\% & NA \\
      \cline{1-3}
   \end{tabularx}
  \end{table}

}

\section{Related Work} \label{sec:related_work}

Prior studies can be broadly classified into two categories: 
(a) techniques for solving larger problems on QAs, which are relevant primarily to Skipper; 
and (b) approaches for enhancing QA fidelity, which are considered related work to both Skipper and Skipper-G.

Prior research on solving larger problems with smaller QAs~\cite{pelofske2022solving, okada2019improving} are iterative schemes, which tend to lose reliability as problem size increases due to reliance on approximations.
Conversely, Skipper explores the entire search space without resorting to approximations.
Recent studies have introduced schemes for addressing larger instances of Boolean Satisfiability (SAT)~\cite{tan2023hyqsat}, Max-Clique~\cite{pelofske2023solving, pelofske2019solving, pelofske2022parallel, pelofske2021decomposition}, and compressive sensing with matrix uncertainty~\cite{ayanzadeh2019quantum} problems. 
However, these methods are specific to their respective applications and are not transferable to other domains, whereas Skipper is versatile and applicable to any application.

Policies for improving the fidelity of QAs can be classified as: 
(a) Preprocessing~\cite{pelofske2019optimizing,ayanzadeh2022equal,ayanzadeh2020reinforcement}, modifying QMIs before submission;
(b) Postprocessing~\cite{ayanzadeh2021multi}, enhancing outcomes using heuristics;
(c) Hybrid strategies~\cite{ayanzadeh2022quantum,ayanzadeh2020leveraging}, combining heuristics and QAs for reliability; 
(d) Logical analog qubits~\cite{jordan2006error,sarovar2013error,young2013error,pudenz2014error,vinci2015quantum,venturelli2015quantum,vinci2016nested,matsuura2016mean,mishra2016performance,matsuura2017quantum,vinci2018scalable,pearson2019analog,matsuura2019nested,mohseni2021error}, spreading qubit information over multiple physical qubits;
and ensembling policies~\cite{ ayanzadeh2022equal,ayanzadeh2020reinforcement,ayanzadeh2020ensemble}, subjecting the quantum program to different noise profiles to suppress the bias. 
These proposals are orthogonal  to Skipper and Skipper-G and can effectively boost the reliability of our proposed techniques.

Skipper is inspired by FrozenQubits~\cite{ayanzadeh2023frozenqubits} in digital QCs. 
While FrozenQubits enhances fidelity of optimization applications in digital QCs, Skipper excels in addressing larger problems and enhancing QA fidelity. 
More importantly, while FrozenQubits' performance diminishes with increased problem graph density, Skipper and Skipper-G maintain their performance, demonstrating the effectiveness of our proposal in handling sparse to dense graphs.

\section{Conclusion}

We propose \emph{Skipper}, a software scheme designed to enhance the capacity and fidelity of QAs. 
Observing that chain lengths in QAs follow a ``Power-Law'' distribution, with a few \emph{dominant chains} containing significantly more qubits than others, Skipper prunes these chains. 
This approach replaces their corresponding program qubits with two possible measurement outcomes, freeing all qubits in the dominant chains and an additional 25\% of isolated qubits previously entrapped in chains. 
Our experiments on a 5761-qubit QA by D-Wave show that Skipper allows QAs to solve problems up to 59\% larger (Avg. 28.3\%) when up to eleven chains are skipped. 
Additionally, by removing five chains, Skipper substantially improves QA fidelity by up to 44.4\% (Avg. 33.1\%). 

The number of chain cuts in Skipper is user-defined; users can trim up to eleven chains, which necessitates running an average of 1024 (and up to 2048) distinct quantum executables. However, 
this may lead to affordability concerns for some users. 
To mitigate this, we introduce \emph{Skipper-G}, a greedy scheme that prioritizes examining sub-spaces more likely to contain the global optimum. 
When up to eleven chains are pruned, Skipper-G runs a maximum of 23 quantum executables. 
Our experiments show that Skipper-G enhances QA fidelity by up to 40.8\% (Avg. 29.2\%), requiring only 11 quantum executable runs for up to five chain cuts, compared to Skipper's 32 runs.



\clearpage
\bibliographystyle{plain}
\bibliography{references}

\end{document}